\documentclass[preprint,showpacs,preprintnumbers,amsmath,amssymb]{revtex4}%
\usepackage{graphicx}
\usepackage{dcolumn}
\usepackage{bm}
\usepackage{amsmath}
\usepackage{amsfonts}
\usepackage{amssymb}%
\providecommand{\U}[1]{\protect\rule{.1in}{.1in}}
\newcommand{\be}{\begin{equation}}
\newcommand{\en}{\end{equation}}
\newcommand{\bea}{\begin{eqnarray}}
\newcommand{\ena}{\end{eqnarray}}
\begin{document}
\title{Generalized Galileon Scenario Inspires Chaotic Inflation}
\author{Mat\'ias L\'opez}
\email{matias.lopez.g@mail.pucv.cl}
\author{Jorge Maggiolo}
\email{jorge.maggiolo.t@mail.pucv.cl} 
\author{Nelson Videla}
\email{nelson.videla@pucv.cl} 
\affiliation{ Instituto de
F\'{\i}sica, Pontificia Universidad Cat\'{o}lica de Valpara\'{\i}so,
Avenida Brasil 2950, Casilla 4059, Valpara\'{\i}so, Chile.}

\author{Pablo Gonz\'alez}
\email{pgonzalezv@uta.cl}
\affiliation{Sede Esmeralda, Universidad de Tarapac\'a, Avda. Luis Emilio Recabarren 2477, Iquique, Chile}

\author{Grigoris Panotopoulos}
\email{grigorios.panotopoulos@tecnico.ulisboa.pt} \affiliation{ Centro de Astrof\'{\i}sica e Gravita{\c c}{\~a}o, Departamento de F{\'i}sica, Instituto Superior T\'ecnico-IST, Universidade de Lisboa-UL, Av. Rovisco Pais, 1049-001 Lisboa, Portugal}

\date{\today}

\begin{abstract}
We study chaotic inflation with a Galileon-like self interaction $G(\phi,X)\Box \phi$, where $G(\phi,X)\propto X^{n}$. General conditions required for successful inflation are deduced and discussed from the background and cosmological perturbations under slow-roll approximation. Interestingly,
it is found that in the regime where the  Galileon term dominates over the standard kinetic term, 
the tensor-to-scalar ratio becomes significantly suppressed in comparison
to the standard expression in General Relativity (GR). Particularly, we find the allowed range in the space of parameters characterizing the chaotic quadratic and quartic inflation models by considering the current observational data of Planck from the $n_{\mathcal{S}}-r$ plane. Finally, we discuss about the issue if the Galileon term is dominant by the end of inflation, this can affect the field oscillation during reheating.
\end{abstract}

\pacs{98.80.Cq}
\maketitle




\section{Introduction}

During the longest part of its lifetime, the universe has undergone a decelerating expansion, being 
dominated first by radiation and then by matter. However, there are two phases of
accelerated expansion in the history of the universe at very early times and late-times as well.
The first accelerating phase corresponds to inflation \cite{starobinsky1,inflation1,inflation2,inflation3}, which is widely accepted as the standard paradigm for describing the physics of the early universe. The first reason is due to the fact that several long-standing puzzles of the hot big-bang model (HBB), such as the horizon, flatness, and monopole problems, find a natural explanation in the framework of inflationary universe. In addition, and perhaps the most intriguing feature of inflation, is that it gives us a causal explanation of the origin of the cosmic microwave background (CMB) temperature anisotropies \cite{Abazajian:2013vfg}, while at the same time it provides us with a mechanism to explain the large-scale structure (LSS) of the universe, since quantum fluctuations during the inflationary era may give rise to the primordial density perturbations \cite{Starobinsky:1979ty,R2,R202,R203,R204,R205}.

The dynamics of inflation can be studied under the so-called the slow-roll approximation (see, e.g. \cite{Lyth:2009zz}). When the slow-roll approximation breaks down inflation ends and the universe enters into the radiation era of standard hot big-bang cosmology. The transition era after the end of inflation, during which the inflaton is converted into the particles that populate the universe later on is called reheating \cite{reh1,reh2}, the physics of which is complicated, highly uncertain, and in addition it cannot be directly probed by observations. One may obtain, however, indirect constraints on reheating according to the following strategy: First we parametrize our ignorance assuming for the fluid a constant equation-of-state $w_{re}$ during reheating, and then we find certain relations between the duration of reheating $N_{re}$ and the reheating temperature $T_{re}$ with $w_{re}$ and inflationary observables \cite{paper1,paper2,paper3}.

The second accelerating phase of the universe corresponds to the current cosmic acceleration supported from type Ia Supernovae data \cite{SN1,SN2}, CMB data \cite{wmap,planck,Aghanim:2018eyx} as well as Baryon Acoustic Oscillation data \cite{BAO1,BAO2}. The most economical $\Lambda$CDM model, which is based on a positive cosmological constant and cold dark matter, suffers from the cosmological constant problem \cite{weinberg}. Large-scale modifications of General Relativity (GR), such as $f(R)$ theories of gravity \cite{modified1,modified2}, Brans-Dicke theory (BD) \cite{BD}, DGP brane model \cite{dgp} and Galileon gravity \cite{galileon}, are capable of explaining the late-time acceleration of the universe without a cosmological constant. For a review on modified gravity and Cosmology see e.g. \cite{review}. In a little known paper in 1974 Horndeski found the most general scalar-tensor theory having second order equations of motion \cite{horndeski}. It turns out \cite{defelice} that Horndeski's theory includes both the canonical scalar field and k-essence \cite{mukhanov2}, while at the same time accommodates $f(R)$ theories, BD theory and galileon gravity \cite{kobayashi,Kobayashi:2010cm,deffayet,charmousis}. The Horndeski theory provides us with a general framework to accounting for the current accelerated expansion of the universe
and the inflationary phase of the very early universe as well. For a review on current status
of Horndeski's theory see e.g. \cite{Kobayashi:2019hrl}. Regarding the observational constraints, specially those coming from measurement of the speed of gravitational waves (GWs) $c_{\textup{GW}}$, 
restrict dramatically Horndeski's theory. The nearly simultaneous detection of gravitational waves
GW170817 and the $\gamma$-ray burst GRB 170817A provides a tight constraint
on $c_{\textup{GW}}$ \cite{TheLIGOScientific:2017qsa,Monitor:2017mdv}
\begin{equation}
-3\times 10^{-15}<c_{\textup{GW}}-1<7\times 10^{-16},
\end{equation}
which basically means that GWs propagate at the speed of light. Then, in order to have $c_{\textup{GW}}=1$ irrespective of the background
cosmological evolution within Horndeski's theory, its Lagrangian is restricted to be \cite{Baker:2017hug,Langlois:2017dyl}
\begin{equation}
\mathcal{L}=f(\phi)\,R+K(\phi,X)-G(\phi,X)\,\Box \phi,\label{Lh}
\end{equation}
leaving this theory constructed only with non-minimally coupling, k-essence, and cubic Galileon
sectors.

Although single-field slow-roll inflation in GR provides us with the best fit to the data \cite{Akrami:2018odb}, considering alternative, non-standard scenarios, are motivated by the fact that certain scalar potentials for the inflaton coming from Particle Physics \cite{Pich:2007vu,Higgs:1964pj,Englert:1964et}, such as the chaotic quadratic or the chaotic quartic ones \cite{Linde:1983gd}, are ruled out by current data. More generally, the monomial potential $V(\phi)=V_0 (\phi/M_{pl})^p$ is ruled out by Planck 2018 data for $p\geq 2$ \cite{Akrami:2018odb}.
For instance, a non-minimal coupling to gravity can save the quartic potential \cite{Tenkanen:2017jih,ref3}. Potential-driven Galileon inflation was studied in \cite{Tsujikawa:2013ila,paperbase} for a Galileon-self coupling of the form $G(\phi,X)=-X/M^3$, bringing
chaotic inflation to be compatible with current observations 
on the tensor-to-scalar ratio available at that time \cite{wmap}. Nevertheless, the latest data from the Keck Array/BICEP2 and Planck collaborations \cite{Ade:2018gkx} constraints robustly the tensor-to-scalar ratio. In this direction, in \cite{Teimoori:2017jzo} it was studied G-inflation with a generalized expression for the Galileon-self coupling given by $G(\phi,X)\propto X^n$ (suggested for the first time in \cite{DeFelice:2011jm}), while in \cite{Herrera:2018ker} the authors proposed the generalization $G(\phi,X)\propto \phi^{\nu}\,X^n$. In both aforementioned works, it was found that the effect of the power $n$ is
to suppresses the tensor-to-scalar ratio, which can be used to
explore the viability of certain scalar potentials for the inflaton, e.g. chaotic one, which is ruled out by current data.

In this way, the main goal of  the present work is to study the viability of chaotic monomial potential
$V(\phi)=V_0 (\phi/M_{pl})^p$ within the G-inflation scenario (\ref{Lh}) with the following expression for the 
Galileon self-interaction
\begin{equation}
G(\phi,X)=\frac{c}{M^{4n-1}}X^n.
\end{equation}
The plan of our work is as follows: In the next section we briefly present the dynamics of Galileon inflation, and we summarize the basic formulas we shall be using. In section \ref{gchaotic} we
apply the general framework presented in previous section to the chaotic potential, then
we discuss our numerical results for the particular cases of chaotic quadratic and quartic potentials, and we finally summarize our work in the fourth section.

\section{G-INFLATION}\label{Ginf}

In this section we give a brief review on the background dynamics and the cosmological perturbations in the framework of G-inflation with a power-law Galileon self-coupling.

\subsection{Background Dynamics}

Our starting point, is the action for the Galileon scenario (\ref{Lh}) with minimal coupling to gravity, i.e. $f(\phi)=\frac{M_{pl}^2}{2}$, which becomes
 \be
S=\int\sqrt{-g}\left( \frac{M_{pl}^2}{2}R+K(\phi, X)-G(\phi, X)\,\Box \phi\right) d^4x
\,.\label{accion}
 \en
Here, $g$ corresponds to  the determinant of 
metric tensor $g_{\mu\nu}$,  $M_{pl}$ is the reduced Planck mass, $R$ denotes  the Ricci
scalar and   $X=-g^{\mu\nu}\partial_{\mu}\phi\partial_{\nu}\phi/2$. The scalar field
is denoted by  $\phi$  and
the functions $K$ and $G$ have an arbitrary dependence on $X$ and
$\phi$.

By assuming a spatially flat Friedmann-Lema\^{i}tre-Robertson-Walker (FLRW) metric
 and a homogeneous scalar field $\phi=\phi(t)$,  then
 the modified Friedmann equations can be written as \cite{kobayashi,Kobayashi:2010cm}
\be
3M_{pl}^2 H^2+K+\dot{\phi}^2(G_{\phi}-K_{X})-3HG_{X}\,\dot{\phi}^3=0,
\,\label{frwa}
 \en
 and
 \be
 -M_{pl}^2\left(2\dot{H}+3H^2\right)+K-\dot{\phi}^2(G_{\phi}+G_{X}\ddot{\phi})=0,
 \,\label{frwb}
 \en
 where $H=\frac{\dot{a}}{a}$ corresponds to Hubble rate and $a$ denotes the
 scale factor. In the following,  we will consider that  the dots denote
differentiation with  respect to cosmic time and the notation
$K_X$ denotes
 $K_X=\partial K/\partial X$, while $K_{XX}$ corresponds to
 $K_{XX}=\partial ^2K/\partial X^2$, and $G_\phi$ means $G_\phi=\partial G/\partial\phi$, etc.

From variation of the action (\ref{accion})  with respect to
the scalar field we have
$$
3\dot{H}G_{X}\,\dot{\phi}^2+\ddot{\phi}\left[3HG_{XX}\,\dot{\phi}^3
-\dot{\phi}^2(G_{\phi X}-K_{XX})+6HG_{X}\,\dot{\phi}-2G_{\phi}+K_{X}\right]+
$$
\be 3HG_{\phi X}\,\dot{\phi}^3+\dot{\phi}^2(9H^2G_{X}-G_{\phi \phi
}+K_{\phi X})-K_{\phi} -3H\dot{\phi}(2G_{\phi }-K_{X})=0.
\,\label{frwc} \en In  the specific  cases  in which the functions
$K=X-V(\phi)$ (with $V(\phi)$ being the effective potential for
the scalar field) and $G=0$, standard single field inflation in the context of General Relativity (GR) is recovered.

In order to study the model of   G-inflation,
for the function $K(\phi, X)$ we choose
\be
K(\phi, X)=X-V(\phi),
\label{K}
 \en
while for the Galileon self-coupling $G(\phi,X)$, following Refs.\cite{Teimoori:2017jzo, DeFelice:2011jm,Herrera:2018ker}, we take a generalized expression
\begin{equation}
G(\phi,X)=\frac{c}{M^{4n-1}}X^n,\label{GSC}
\end{equation}
where $c$ is a dimensionless constant to be fixed, $M$ is a mass scale, and $n$ is a positive integer power. The case $n=1$ was studied previously in Ref.\cite{paperbase} for chaotic inflation, whose theoretical predictions were consistent with data available at that time.
 
Following Ref.\cite{kobayashi}, we will consider the model
of G-inflation under the slow-roll approximation. In this sense,
the effective potential dominates over the functions $X$,
$|G_{X}H\dot{\phi} ^3|$. Thus,  under this
approach, the Friedmann equation
 given by Eq.(\ref{frwa}) can be approximated to
  \be
 3M_{pl}^2H^2\simeq V(\phi).
 \,\label{frwa2}
 \en
In context of slow-roll approximation, we can
 introduce  the set of slow-roll parameters for G-inflation, defined as \cite{kobayashi}
$$
\delta_X={K_{X} X\over M_{pl}^2H^2},
\quad \delta_{GX}={G_{X} \dot{\phi}X\over M_{pl}^2 H},
$$
\be
\varepsilon_1=-{\dot{H}\over H^2},
\quad \epsilon_2=-{\ddot{\phi}\over H\dot{\phi}}=-\delta_{\phi}.
\label{srp} \en
From the  parameters defined above and combining
 with the Friedmann equations (\ref{frwa}) and
(\ref{frwb}), the slow-roll parameter $\varepsilon_1$ can be
rewritten as \be
 \varepsilon_1=\delta_X+3 \delta_{GX}-\delta_{\phi} \delta_{G
 X}.
\,\label{epsilon1} \en Now,  from the functions $K(\phi,X)$ and
$G(\phi,X)$ given by Eqs.(\ref{K}) and (\ref{GSC}), respectively, and considering the slow-roll
parameters from Eqs.(\ref{srp}) and (\ref{epsilon1}), the equation
of motion for the scalar field is rewritten as
\begin{equation}
3H\dot{\phi}(1-\epsilon_2/3)+
\frac{3n c}{M^{4n-1}}X^{n-1}H^2\dot{\phi}^2(3-\varepsilon_1-2n\epsilon_2)=-V_{,\phi}.\label{frwc2a}
\end{equation}
Within the slow-roll analysis,  we are going to consider that the
slow-roll parameters $|\varepsilon_1|, |\epsilon_2| \ll 1$, see Ref.\cite{kobayashi}. Then, a leading order of slow-roll approximation, the equation of motion for the scalar field, given by
Eq.(\ref{frwc2a}), yields
 \be 3H\dot{\phi}(1+\mathcal{A})
\simeq-V_{,\phi}\,, \,\label{frwc2} \en 
where $\mathcal{A}$ being a function defined as follows
\begin{equation}
\mathcal{A}\equiv 3\frac{\delta_{GX}}{\delta_X}=\frac{3\,c\,n}{2^{n-1}M^{4n-1}}\dot{\phi}^{2n-1}H.\label{AG}
\end{equation}
By combining Eqs.(\ref{frwa2}) and (\ref{frwc2}), the slow-roll parameter $\delta_X$ may bew rewritten as
\begin{equation}
\delta_X \simeq \frac{\epsilon}{(1+\mathcal{A})^2},
\end{equation}
where $\epsilon=\frac{M_{pl}^2}{2}\left(\frac{V_{,\phi}}{V}\right)^2$ is the usual slow-roll
parameter for standard inflation. Accordingly, the slow-roll parameter $\varepsilon_1$ now becomes
\begin{equation}
\varepsilon_1=-\frac{\dot{H}}{H^2}\simeq (1+\mathcal{A})\delta_X\simeq \frac{\epsilon}{1+\mathcal{A}}.\label{varepsilon}
\end{equation}
As it can bee seen from Eq.(\ref{varepsilon}), the conventional slow-roll inflation corresponds to the limit $\mathcal{A}\rightarrow 0$, in which $\varepsilon_1\simeq \epsilon \simeq \delta_X$. For small 
$M$, there appears a regime where the Galileon self-interaction dominates over the standard kinetic
during inflation, i.e. $\mathcal{A}\gg 1$ ($\delta_{GX}\gg \delta_X$), and the evolution of $\phi$ slows down relative to those in standard inflation.

An important issue is the appearance of ghosts and Laplacian instabilities in the regime $\mathcal{A}\gg 1$ (see , e.g. Ref.\cite{DeFelice:2010nf} for an extensive analysis). From Eq.(\ref{AG}), and noting 
that, during inflation, $V_{,\phi}>0$ and $\dot{\phi}<0$, in order to avoid the appearance of ghosts and Laplacian instabilities, we demand the condition $c\,\dot{\phi}^{2n-1}>0$ ($c=-1$). From Eq.(\ref{varepsilon}), the end of inflation is now determined by 
\begin{equation}
\epsilon_V(\phi_{end})=1+\mathcal{A}(\phi_{end})\label{phiend}.
\end{equation}
The number of $e$-folds in the slow-roll approximation we obtain
\begin{equation}
{\mathcal{N}}\simeq \frac{1}{M_{pl}^2}\int_{\phi_{end}}^{\phi_{*}}\,(1+\mathcal{A})\frac{V}{V_{,\tilde{\phi}}}\,d\tilde{\phi},\label{NN}
\end{equation}
where $\phi_{*}$ and $\phi_{end}$ are the values of the scalar field when the cosmological scales crosses the Hubble-radius and at the end of inflation, respectively.
As it can be seen, the number of $e$-folds is enhanced due to an extra term
of $(1+{\mathcal{A}})$. This implies a more amount of inflation, between these two values of the field, compared to standard inflation.

\subsection{Perturbations}

In the following, we present a brief review of the basic relations governing the dynamics of
cosmological perturbations in the framework of G-inflation, based mainly on
Refs.\cite{kobayashi,DeFelice:2011uc,DeFelice:2013ar}. 

Regarding the power spectrum of the primordial scalar
perturbations $\mathcal{P}_{\mathcal{S}}$, in the slow-roll
 approximation it can be written as
 \be
\mathcal{P}_{\mathcal{S}}\simeq \left.\frac{H^2\,q_s^{1/2}}{8\pi M_{pl}^2\,\varepsilon_{\mathcal{S}}^{3/2}}\right|_{c_s k=aH} \,,\label{PR}
\en 
where the functions $q_s$ and $\varepsilon_{\mathcal{S}}$ are defined as
\be q_s=M_{pl}^2\left(\delta_X+2\delta_{XX}+6 \delta_{GX}+6 \delta_{GXX}-2
\delta_{G\phi} \right)\,,\label{qs} \en and \be \varepsilon_{\mathcal{S}}=\delta_X+4
\delta_{GX}-2 \delta_{G\phi}
\,,\,\,\,\,\mbox{where}\,\,\,\,\delta_{XX}={K_{XX} X^2\over M_{pl}^2 H^2},
\quad\mbox{and}\,\,\,\, \delta_{GXX}={G_{XX} \dot{\phi}X^2\over
M_{pl}^2 H}.\label{ves} \en
 Here, $c_s^2$ is the propagation speed of a scalar mode squared, which is defined through the
 relation
\begin{equation}
c_s^2={\varepsilon_{\mathcal{S}}\over q_s}M_{pl}^2.\label{cs2}
\end{equation}
In this form, by assuming the explicit form of functions $K$ and $G$, given respectively by Eqs.(\ref{K}) and
(\ref{GSC}), it is found that the
functions $q_s$ and $\varepsilon_{\mathcal{S}}$  are rewritten as
\begin{equation}
q_s={X\over H^2}\left( 1+2n\mathcal{A}\right),
\,\,\,\mbox{and}\,\,\,\,\,\,\varepsilon_{\mathcal{S}}={X\over M_{pl}^2 H^2}\left(
1+{4\over 3}\mathcal{A}
\right).\label{qses}
\end{equation}
From Eq.(\ref{PR}) and considering functions already defined above, the
scalar power spectrum in the slow-roll approximation results
\be
\mathcal{P}_{\mathcal{S}}\simeq{H^2(1+2n\mathcal{A})^{1/2}\over8\pi^2
M_{pl}^2 \delta_X(1+4\mathcal{A}/3)^{3/2}}
\simeq{V^3(1+\mathcal{A})^{2}(1+2n\mathcal{A})^{1/2}\over 12\pi^2 M_{pl}^6
    V_{\phi}^2(1+4\mathcal{A}/3)^{3/2}}\,,\label{PR2} \en and  the scalar
propagation speed squared becomes  
\begin{equation}
 c_s^2={1+4\mathcal{A}/3\over
1+2n\mathcal{A}} \leq1.\label{cs2n}
\end{equation}
In particular, in the limit $\mathcal{A}\gg1$, $c_s^2$ reduces to
\begin{equation}
c_s^2\simeq \frac{2}{3\,n},\label{csA}
\end{equation}
where the power $n$ is such that
$n\geq2/3$. Then, the background dynamics evolves such that
Eqs.(\ref{qses}) and (\ref{cs2n}) yield $q_s>0$ and $c_s^2>0$, avoiding
Laplacian instabilities and ghosts.

In the limit ${\cal{A}}\gg1$, the scalar power
spectrum, given by Eq.(\ref{PR2}), becomes approximately
\begin{equation}
\mathcal{P}_{\mathcal{S}}\simeq{3H^4\sqrt{6n}\over64\pi^2
X\mathcal{A}}\simeq{\sqrt{6n}\,V^3 \mathcal{A}\over 32\pi^2
M_{pl}^6 V_{\phi}^2}.\label{P1}
\end{equation}
Also, the scalar spectral index $n_{\mathcal{S}}$ associated with the tilt of
the power spectrum, characterizes its scale dependence and it 
is defined as  $n_{\mathcal{S}}-1=\left.\frac{d\ln{\cal{P_S}}}{d\ln k}\right|_{c_s k=aH}$.
Thus, from Eq. (\ref{PR2}) and considering that
under slow-roll approximation $d \ln k\simeq Hdt$, the scalar spectral index yields
\begin{equation}
n_{\mathcal{S}}\simeq\,1-\frac{6\epsilon}{1+{\cal{A}}}+\frac{2\eta}{1+\cal{A}}
+{\dot{\cal{A}}\over
H}\left[\frac{2}{1+{\cal{A}}}+\frac{n}{1+2n{\cal{A}}}
-\frac{2}{1+4{\cal{A}}/3}\right],\label{ns0}
\end{equation}
where $\epsilon$ and $\eta$ are the standard slow-roll parameters,
defined as
\begin{equation}
  \epsilon=\frac{M_{pl}^2}{2}\left(\frac{V_{\phi}}{V}\right)^2,\;\,\,
\,\,\,\mbox{and}\,\,\,\,\,\,\eta=M_{pl}^2\,\frac{V_{\phi\phi}}{V},
\end{equation}
respectively. Here, we observe that in the limit
${\cal{A}}\rightarrow 0$, the
scalar spectral index given by Eq.(\ref{ns0}) coincides with the
 expression obtained in standard inflation in GR, where  $n_{\mathcal{S}}\simeq 1-6\epsilon+2\eta$. On
 the other hand, in the
limit $\mathcal{A}\gg 1$, where the Galileon term dominates the
inflaton dynamics, the  scalar index $n_{\mathcal{S}}$ results
\begin{equation}
n_{\mathcal{S}}\simeq 1-\left(5+\frac{1}{n}\right)\frac{\epsilon}{\mathcal{A}}+\left(1+\frac{1}{2n}\right)\frac{\eta}{\mathcal{A}}\,\,\,.\label{ns}
\end{equation}

Regarding tensor perturbations in the framework of
 G-inflation, the expression for the power spectrum becomes similar to  
 those obtained in standard inflation in GR
\begin{equation}
{\cal{P}}_{\mathcal{T}}={2H^2\over \pi^2 M_{pl}^2 }.
\end{equation} 
Accordingly, the tensor-to-scalar ratio, defined as $r={\cal{P}}_{\mathcal{T}}/\mathcal{P}_{\mathcal{S}}$, in the
framework of G-inflation under slow-roll approximation can be
written as
\begin{equation}
r=\frac{{\cal{P}}_{\mathcal{T}}}{{\cal{P}}_{\cal{S}}}\simeq\,16\epsilon\,\left[\frac{(1+4{\cal{A}}/3)^{3/2}}
{(1+{\cal{A}})^2(1+2n{\cal{A}})^{1/2}}\right].\label{rl}
\end{equation}
Again, we note that in the limit ${\cal{A}}\rightarrow 0$, the
tensor-to-scalar  ratio  coincides with the expression obtained in
standard inflation, where $r\simeq16\epsilon$. Now, in the
opposite limit, $ {\cal{A}}\gg 1$, the tensor-to-scalar ratio $r$ is
approximated to
\begin{equation}
r\simeq\frac{4\sqrt{2}}{3^{3/2}}\,\frac{16\epsilon}{\sqrt{n}\,\,{\cal{A}}},\label{r}
\end{equation}
which is suppressed by a factor $\sim \sqrt{n}\,\mathcal{A}$ in comparison
to the standard expression in GR. Moreover, the expression of
above agrees with Ref.\cite{paperbase} for $n=1$. With is, Galileon inflation becomes phenomenologically  distinguishable  from  standard  inflation, which enables us to
explore the viability of certain scalar potentials for the inflaton, such as monomial one, which is ruled out by current data.

\section{G-inflation with a chaotic potential: results for $\mathcal{A}\gg 1$}\label{gchaotic}

For concreteness and comparison with previous works, we are going to study chaotic inflation, characterized by a potential of the form 
\begin{equation}
V(\phi)=V_0 (\phi/M_{pl})^p\quad , \quad (p>0),\label{Vphi}
\end{equation}
where $V_0$ and $p$ are real constants and $M_{pl}$ is the reduced mass Planck. In addition, we shall consider the presence of the Galileon-like self-interaction given by Eq.(\ref{GSC}). In order to derive analytical expressions for the quantities which describe the background dynamics, we restrict ourselves to the regime dominated by the Galileon term, ${\mathcal{A}}\gg 1\,(M\rightarrow 0)$. Otherwise, the full analysis of the model requires numerical solving of the background as well as perturbation dynamics.

For the background dynamics of our concrete model, it is found that
$\mathcal{A}$ as a function of the inflaton fields is given by
\begin{equation}
{\mathcal{A}}(\phi)=6^{\frac{1}{2}\left(\frac{1}{n}-1\right)}n^{\frac{1}{2n}}p^{1-\frac{1}{2n}}\frac{\sqrt{V_0}}{M^2}\left(\frac{M}{M_{pl}}\right)^{\frac{1}{2n}} \left(\frac{\phi}{M_{pl}}\right)^{\frac{p}{2}+\frac{1}{2n}-1}.\label{Aphi}
\end{equation}
After replacing the equation of above into Eq.(\ref{phiend}), we may compute the value of the scalar field at the end of inflation $\phi_{end}$, yielding
\begin{equation}
\frac{\phi_{end}}{M_{pl}}=\left[\frac{3^{n-1}p^{2n+1}}{2^{n+1}n}\frac{M_{pl}}{M}\left(\frac{M^{4n}}{V_0^2}\right)\right]^{\frac{1}{1+n(2+p)}}.\label{phiendc}
\end{equation}
Then, analytical integration of Eq.(\ref{NN}) gives us an
expression for the scalar field at the Hubble-radius crossing $\phi_*$ in terms of $\mathcal{N}$ and 
the parameters characterizing our model.

Substituting the previous solution (not shown) into Eq.(\ref{P1}) and using the Planck normalization
$\mathcal{P}_{\mathcal{S}}=2.169\times 10^{-9}$ \cite{Akrami:2018odb}, we find
\begin{equation}
\frac{V_0}{M_{pl}^{4}}\left(\frac{M_{pl}}{M}\right)^{\frac{p(1-4n)}{2n+1}}=\left[\alpha_0 \frac{\left(2.169\times 10^{-9} \pi^2 p\right)^{1+n(2+p)}}{\left(\frac{\mathcal{N}}{3}\left[1+n(2+p)\right]+\frac{np}{3}\right)^{1+n(2+3p)}}\right]^{\frac{1}{2n+1}},\label{PSN}
\end{equation}
where $\alpha_0$ is a constant defined as
\begin{equation}
\alpha_0=\left(\frac{1}{9 p}\right)^{p}2^{11+22n+2p+13np}\,n^{\frac{1+2n+2p+5np}{2}}.
\end{equation}
It is worth to mention that Eq.(\ref{PSN}) evaluated at $n=1$ reduces to those found on Ref.\cite{paperbase}. If the power $n$ of the generalized expression of the Galileon-self coupling has a fixed value, $V_0$ tends to be larger for smaller $M$. Now, when the mass scale $M$ is fixed, $V_0$ tends to be smaller for larger $n$.

The predicted scalar spectral index (\ref{ns}) and the tensor-to-scalar ratio $r$ (\ref{r}), both expressed in terms of the number of $e$-folds $\mathcal{N}$, become
\begin{eqnarray}
n_{\mathcal{S}}&=&1-\frac{1+(2+3p)n}{\left(1+n(2+p)\right){\mathcal{N}}+np },\label{nsN}\\
r&=&\frac{64 \sqrt{6}}{9}\frac{p \sqrt{n}}{\left(1+n(2+p)\right){\mathcal{N}}+np }.\label{rN}
\end{eqnarray}

Again, these expressions reduced to those obtained in Ref.\cite{paperbase} for $n=1$. Before study some particular cases, we way analyse the effect of the power $n$ on the values for $n_{\mathcal{S}}$ and $r$. In particular, for larger $n$, the scalar spectral index tends to $\frac{(2+p){\mathcal{N}}-2(p+1)}{(2+p){\mathcal{N}}+p}$, while the tensor-to-scalar ratio tends to zero. For concreteness, we shall study the quadratic potential ($p=2$) and the quartic one ($p=4$) separately. 

\subsection{$p=2$}

\begin{figure}[ht!]
\centering
\hspace{-1 in}
\includegraphics[scale=0.3]{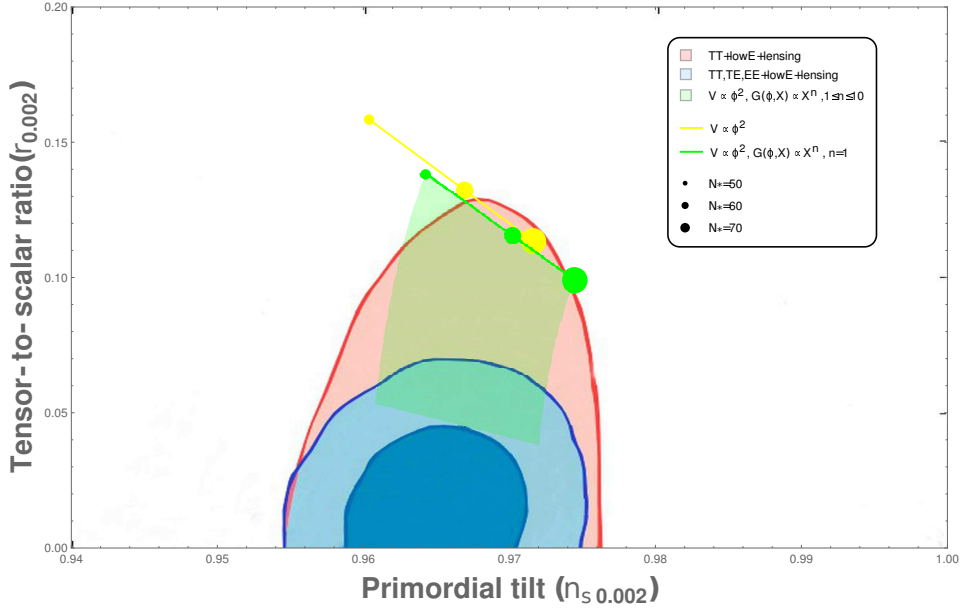}
{\vspace{0 in}}
\caption{Allowed contours at the 68 and 95 $\%$ C.L., from the latest Planck data \cite{Akrami:2018odb} and theoretical predictions in the $n_{\mathcal{S}}-r$ plane for chaotic quadratic inflation ($p=2$)
in our model (green-shadow region) and the standard scenario (yellow-line).}
\label{fig1} 	
\end{figure}

In first place, we shall focus on chaotic quadratic inflation ($p=2$). Then, 
$n_{\mathcal{S}}$ (\ref{nsN}) and $r$ (\ref{rN}) are certain functions of
the number of $e$-folds $\mathcal{N}$ and the power $n$. In
this way, we plot parametrically $r$ versus $n_{\mathcal{S}}$, by varying 
$\mathcal{N}$ and $n$ simultaneously in a wide range, in the same plot with the
allowed contour plots of the latest Planck data as well the theoretical predictions
for standard chaotic quadratic inflation (yellow line), as is shown in Fig.\ref{fig1}. On the $n_{\mathcal{S}}-r$ plane, as $n$ increases for a fixed $\mathcal{N}$ the shown curves lead
to lower tensor-to-scalar ratio. Hence, the theoretical
prediction lies inside the 95 $\%$ C.L. region from Planck 2018 \cite{Akrami:2018odb}
when the power $n$ takes the following values:
\begin{itemize}
\item For $\mathcal{N}=50$,
\begin{equation}
n\gtrsim 8.
\end{equation}
\item For $\mathcal{N}=60$,
\begin{equation}
n\gtrsim 4.
\end{equation}
\item For $\mathcal{N}=70$,
\begin{equation}
n\gtrsim 5.
\end{equation}
\end{itemize}

As it can be seen from Fig.\ref{fig1}, for $n=1$ the tensor to-scalar ratio becomes slowly decreased in comparison to the standard scenario, e.g., for $\mathcal{N}=60$ we have that 
$r\simeq 0.115$, which agrees with Ref.\cite{paperbase}. Nevertheless, this former result was supported by last data of WMAP \cite{wmap}, the current available data at that time. Otherwise, for $n\simeq 4$, the scalar spectral index and tensor-to-scalar ratio becomes $n_{\mathcal{S}}\simeq 0.968$
and $r\simeq 0.07$ at $\mathcal{N}=60$, being supported by current bounds of latest Planck data.

Considering that the lower bound on $n$ for $\mathcal{N}=60$ is $n\simeq4$, Eq.(\ref{Aphi}), leads to the following relation 
\begin{equation}
\mathcal{A}\simeq 3.23 \times 10^{-10} \left(\frac{M_{pl}}{M}\right)^{10/3},
\end{equation}
which means that, in order to be within the regime $\mathcal{A}\gg 1$, the mass scale is such that
$M\ll 1.42 \times 10^{-3}\,M_{pl}$. Accordingly, the Planck normalization (\ref{PSN}) yields 
$V_0\gg 8.49 \times 10^{-12}\,M_{pl}^4$. Now, From the expression for the inflaton potential (\ref{Vphi}), if we identifying $V_0\sim m^2\,M_{pl}^2$, the former constraint on $V_0$ translates into a lower bound for the mass of the inflaton given by $m\gg 2.91 \times 10^{-6}\,M_{pl}^2$. In this context, the mass for the inflaton field can be even larger than those predicted in standard inflation.

\subsection{$p=4$}

\begin{figure}[ht!]
\centering
\hspace{-1 in}
\includegraphics[scale=0.3]{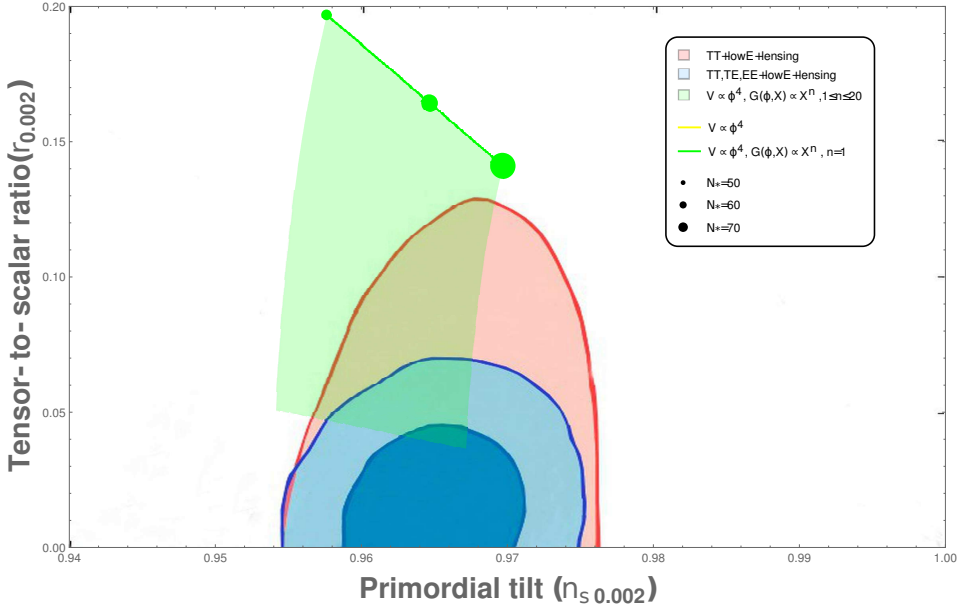}
{\vspace{0 in}}
\caption{Allowed contours at the 68 and 95 $\%$ C.L., from the latest Planck data \cite{Akrami:2018odb} and theoretical predictions in the $n_{\mathcal{S}}-r$ plane for chaotic quartic inflation ($p=2$)
in our model (green-shadow region).}
\label{fig2} 	
\end{figure}

Now, we turn on the theoretical predictions for the particular case of chaotic quartic inflation ($p=4$). In similar fashion as we did for the chaotic quadratic potential, we plot parametrically $r$ (\ref{rN}) versus $n_{\mathcal{S}}$ (\ref{nsN}),  by varying 
$\mathcal{N}$ and $n$ simultaneously in a wide range, in the same plot with the
allowed contour plots of the latest Planck data, as is shown in Fig.\ref{fig2}. In this case
, the theoretical
prediction of standard chaotic quartic inflation is not 
shown explicitly, since the value of the tensor-to-scalar ratio $r$ is well outside
the range constrained by Planck 2018 data. As it can be seen from Fig.\ref{fig2}, as 
the power $n$ increases for a fixed $\mathcal{N}$, the shown curves lead
to lower tensor-to-scalar ratio. Moreover, the curve for $\mathcal{N}=50$ is always outside
the 95 $\%$ C.L. region for any value of $n$. Otherwise, for $\mathcal{N}=60$ and $\mathcal{N}=70$, the theoretical
predictions enter to the 95 $\%$ C.L. region from Planck 2018
when the power $n$ takes the following values:
\begin{itemize}
\item For $\mathcal{N}=60$,
\begin{equation}
n\gtrsim 9.
\end{equation}
\item For $\mathcal{N}=70$,
\begin{equation}
n\gtrsim 6.
\end{equation}
\end{itemize}

For a sake of comparison, the tensor-to-scalar ratio for $n=1$ at $\mathcal{N}=60$ is $r\simeq 0.164$, which agrees with those found in Ref.\cite{paperbase}, being also supported
by current observational data at that time \cite{wmap}, but incompatible with current Planck data. Recall that, in order to be in agreement with the current upper bound on the tensor-to-scalar ratio, the power $n$ of the Galileon coupling
must satisfy the lower bound $n\gtrsim 9$, for $\mathcal{N}=60$. Particularly, for $n\simeq 9$, the scalar spectral index and tensor-to-scalar ratio becomes $n_{\mathcal{S}}\simeq 0.962$
and $r\simeq 0.06$, being supported by current bounds of latest Planck data.

With regard to the consistency of the dynamics evolving according to the 
Galileon dominated regime $\mathcal{A}\gg1$, from Eq.(\ref{Aphi}) it is found that the mass scale $M$ for $\mathcal{N}=60$ satisfies $M\ll 1.61\times 10^{-3}\,M_{pl}$. On the other hand, by
identifying $V_0\sim \lambda\,M_{pl}^4$, Planck normalization
(\ref{PSN}) and the lower limit for $M$ set a lower bound for $\lambda$, yielding $\lambda\gg 3.67 \times 10^{-15}$. Hence, the prediction for the coupling $\lambda$ within
standard inflation becomes smaller than those we already found in our generalized Galileon scenario.

\subsection{The issue of instabilities when $\mathcal{A}\gg 1$}

As we have seen, for both chaotic quadratic and quartic potential, the tensor-to-scalar ratio
$r$ gets smaller, yielding that the theoretical curves lie inside the $95\%$ C.L. as well as
$68\%$ C.L. observational contours. Nevertheless, if the Galileon
self-interaction is still dominating over the standard kinetic term after the end of inflation, the coherent oscillations of the inflaton field are spoiled \cite{paperbase}. Roughly speaking, $\dot{\phi}$
passes from $\dot{\phi}>0$ to $\dot{\phi}<0$ , which translates into a negative propagation
speed squared of a scalar mode, $c_s^2<0$, leading in turn to the instability of small-scale
perturbations. 

A possible way out of the issue of above is to study the dynamics
of our scenario in a full regime, without any approximation. As a first approach, we solve
numerically the full background equations of motion (\ref{frwa}), (\ref{frwb}), and (\ref{frwc}) from inflation up to
the oscillatory regime. We restrict ourselves to the case $p=2$, and as a first approach, the parameter
values characterizing the model are assumed that do not differ from those already obtained in the regime $\mathcal{A}\gg 1$. 

\begin{figure}[ht!]
\centering
{\hspace{-0.55 in}}
\includegraphics[scale=1.5]{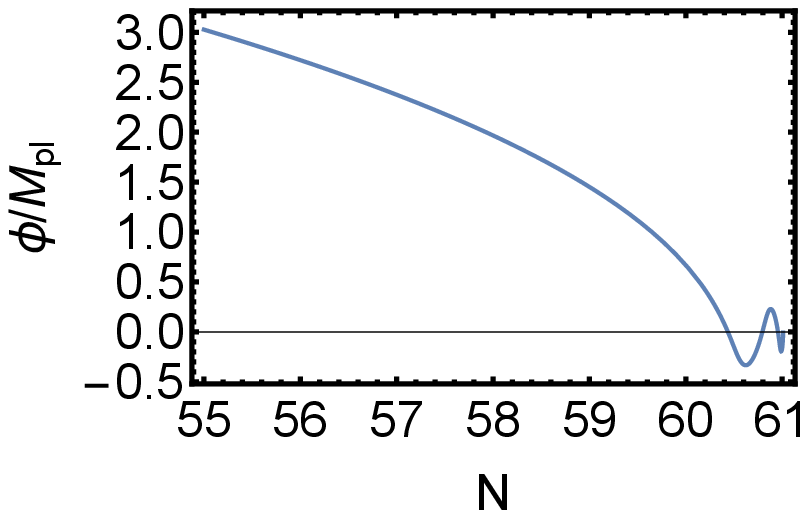}
\includegraphics[scale=1.4]{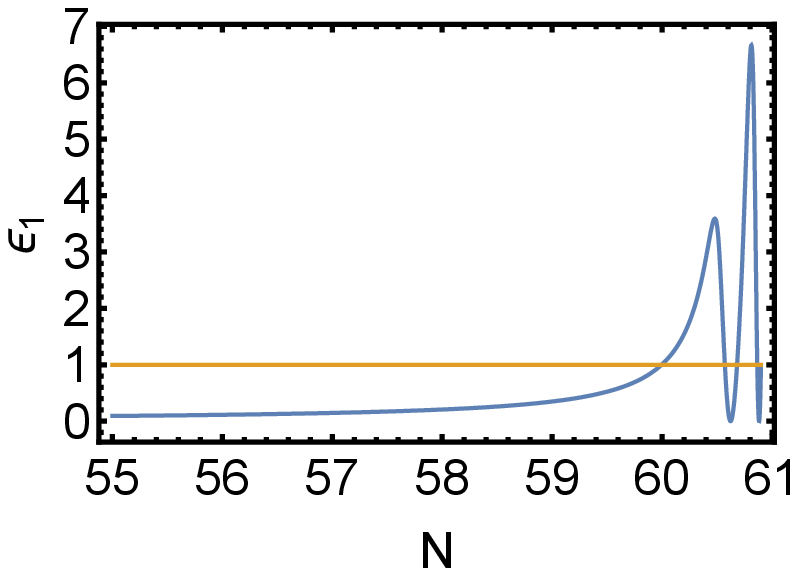}
{\vspace{0 in}}
\caption{Upper and lower plot depict the evolution of slow-roll parameter $\varepsilon_1$ and
the inflaton field from the last 5 $e$-folds on inflation and the end of oscillatory stage, respectively.}
\label{fig3} 	
\end{figure}

\begin{figure}[ht!]
\centering
\includegraphics[scale=1.4]{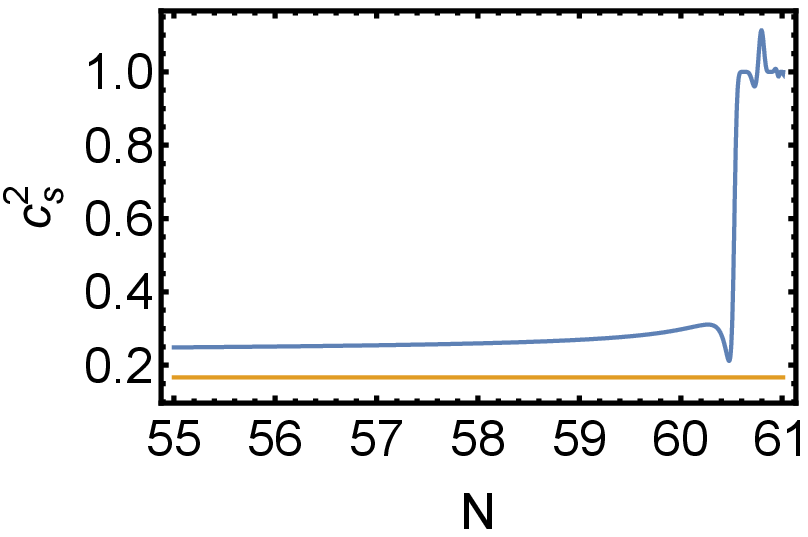}
\includegraphics[scale=1.4]{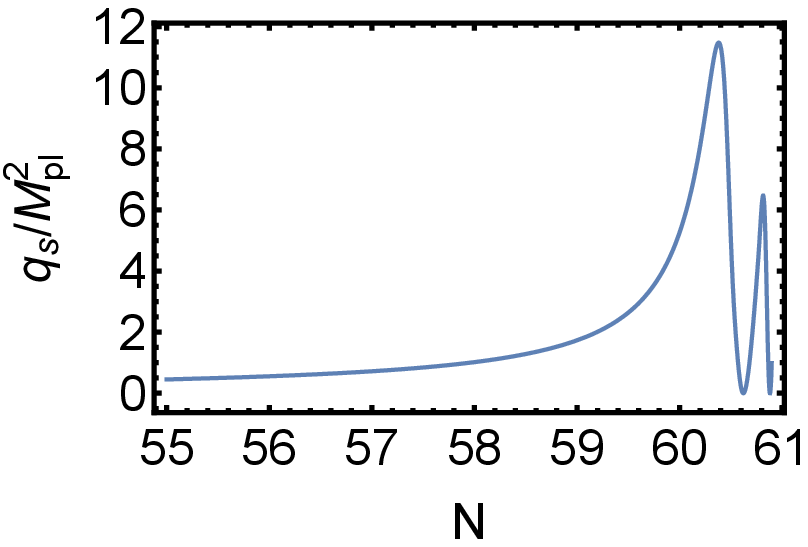}
{\vspace{0 in}}
\caption{Upper and lower plot depict the evolution of the scalar propagation speed squared $c_s^2$ and the function $q_s$ during inflation as well as
during the oscillatory stage, respectively.}
\label{fig4} 	
\end{figure}

The numerical procedure is summarized as follows: we rewrite Eqs.(\ref{frwa}), (\ref{frwb}), and (\ref{frwc})
in terms of the the number of $e$-folds $N$, which relates to the Hubble rate $H$ through
$dN=Hdt$. Recall that Eqs.(\ref{frwa}), (\ref{frwb}), and (\ref{frwc}) are not independent, hence
me may solve a system of two coupled differential equations for $\phi(N)$ and $H(N)$. Considering
that slow-roll is an attractor, the initial conditions for $\phi$, $\dot{\phi}$, and $H$ can be derived from the slow-roll equations themselves. Then, after solving for $\phi(N)$ and $H(N)$, we compute
the slow-roll parameter $\varepsilon_1$, the scalar propagation speed squared $c_s^2$ as well
as the function $q_s$. In order to get 60 $e$-folds of inflation and the behaviour of the solutions
shown in Figs.(\ref{fig3}) and (\ref{fig4}), we choose the following set of values for the specific chaotic quadratic potential ($p=2$) model:
\begin{equation}
n=4,\quad M=1.1\times 10^{-3}\,M_{pl},\quad V_0=2.02\times 10^{-11}\,M_{pl}.\label{values}
\end{equation}

The upper plot of Fig.(\ref{fig3}) shows the evolution of inflaton field for the last
5 $e$-folds of inflation and the subsequent oscillatory stage. As it can seen, the transition from inflation and the time when damped oscillations take place is almost instantaneous. In practice, the duration of last stage is around one $e$-fold before the numerical computation stops. The lower 
plot depicts the evolution of the slow-roll parameter $\varepsilon_1$ (\ref{epsilon1}) for the last
5 $e$-folds of inflation and the subsequent oscillatory stage. Note that during slow-roll inflation $\varepsilon_1\ll1$ and at the end of inflation it becomes equal to one at $N=60$. Shorty after the end of inflation takes place the kinetic epoch and then, $\varepsilon_1$ becomes to oscillate during 
around one $e$-fold before the numerical computation stop. Since there is no coupling of
the inflaton to other matter components, e.g. a radiation fluid, $\varepsilon_1$ will oscillate and does not stabilize to a fixed value (see Ref.\cite{Gonzalez:2018jax} for a further analytical and numerical analysis).

The behaviour of the scalar propagation speed squared $c_s^2$ during inflation as well as
during the oscillatory stage is shown in upper plot of Fig.(\ref{fig4}). In the same plot
we compare the full expression for $c_s^2$ (blue line) with those obtained under the slow-roll approximation within the regime $\mathcal{A}\gg1$, $c_s^2=1/6$ (by evaluating Eq.(\ref{csA}) at $n=4$). As it can be seen, the full expression for $c_s^2$ remains
constant and positive during the whole inflationary stage, then right after inflation it is
suppressed and then rapidly becomes equal to one, and stars to oscillate during around  one $e$-fold before the computation stops. In addition, lower plot shows how the function
$q_s$ evolves during inflation as well as during the oscillatory phase. In particular, $q_s$ increases during the last $e$-folds of inflation, reaching a maximum value shortly after the
end of inflation and then starts to oscillate, always taking positive values. In this way, we find
that the inflaton oscillates if the mass scale is such that
\begin{equation}
M\gtrsim 1.1\times 10^{-3}\,M_{pl},
\end{equation}
which in turn provides that the conditions $c_s^2>0$ and $q_s>0$ are satisfied during and after inflation. It is worth mentioning that the above results is within the range obtained for chaotic 
quadratic inflation for $n=1$ in Ref.\cite{paperbase}. At this point we would like to stress
that previous analysis 
is intended to be a first approach to overcome the issue of instabilities. In this direction, a further
numerical treatment of background as well as perturbative dynamics is beyond the scope of the present work. 

\section{Conclusions}

To summarize, in the present work we studied the viability of chaotic potentials within
G-inflation scenario, where the Galileon self-coupling has a power-law form $G(\phi,X)=\frac{c}{M^{4n-1}}X^n$, with $M$ and $n$ being a mass scale and a positive integer power and $M$, respectively. Firstly, we developed the theoretical framework of potential driven inflation with
this generalized Galileon self-coupling at background as well as perturbative levels under the slow-roll approximation. In particular, we derive the expression for the observables as
the scalar power spectrum, scalar spectral index and the tensor-to-scalar ratio. Interestingly,
it was found that in the regime where the  Galileon term dominates over the standard kinetic term $\mathcal{A}\gg1$,
the tensor-to-scalar ratio becomes significantly suppressed by a factor $\sim \sqrt{n}\,\mathcal{A}$ in comparison
to the standard expression in GR. This distinguishability, at phenomenological level, enabled us the explore the viability of certain scalar potentials for the inflaton, such as monomial one $V(\phi)=V_0 (\phi/M_{pl})^p$, which is ruled out by current data for $p\geq 2$. Accordingly, for this monomial potential, we analysed the effect of the power $n$ on the values for $n_{\mathcal{S}}$ and $r$. In particular, for larger $n$, the scalar spectral index tends to $\frac{(2+p){\mathcal{N}}-2(p+1)}{(2+p){\mathcal{N}}+p}$, while the tensor-to-scalar ratio tends to zero. As a specific examples, we studied the cases of chaotic quadratic ($p=2$) and
quartic quartic ($p=4$). For cases, in order to obtain analytic expressions for the background quantities and observables as the scalar power spectrum, scalar spectral index and the tensor-to-scalar ratio as functions of the number of $e$-folds, we restrict ourselves to the 
regime where the  Galileon term dominates over the standard kinetic term. For each case, by means the current observational bounds on the inflationary observables, we found constraints on the power $n$, the amplitude of the potential $V_0$, and the mass scale $M$ which
characterizes the Galileon self-coupling. In this way, monomial potential is bring to be
compatible with current observations in this generalized Galileon scenario for $\mathcal{A}\gg1$. However, the issue of instabilities arises if the Galileon term is still dominating over the standard kinetic term after the end of inflation, leading to a negative propagation
speed squared of a scalar mode, $c_s^2<0$, and subsequently to the instability of small-scale
perturbations. In order to clarify this issue, as first approach, we restrict to solve numerically the full background equations
for the chaotic quadratic ($p=2$) potential and it was found that, in order to avoid
the appearance of ghosts and Laplacian instabilities during the subsequent post-inflationary stage, the mass scale $M$ must satisfy the condition $M\gtrsim 1.1\times 10^{-3}\,M_{pl}$. This ensures
the coherent oscillations of the inflaton field during reheating and the transition
to standard radiation-dominated era of Hot Big-Bang. In this direction, 
a more detailed analysis of the post-inflationary phase should be performed
in order to obtain additional constraints on this class of models, particularly from 
the duration of reheating $N_{re}$ and the reheating temperature $T_{re}$. We hope to be able to address this point in a future work.


\section*{Acknowlegements}

The author N.~V. was supported by Comisi\'on Nacional de Ciencias y Tecnolog\'ia of Chile through FONDECYT Grant N$^{\textup{o}}$ 11170162.
The author G.~P. thanks the Funda\c c\~ao para a Ci\^encia e Tecnologia (FCT), Portugal, for the financial support to the Center for Astrophysics and Gravitation-CENTRA, Instituto Superior T{\'e}cnico,  Universidade de Lisboa, through the Grant No. UID/FIS/00099/2013. He also thanks the Pontificia Universidad Cat\'{o}lica de Valpara\'{\i}so, where part of the work was completed, for its warmest hospitality.



\begin{thebibliography}{99}
\bibitem{starobinsky1} A.~A.~Starobinsky,
  Phys.\ Lett.\  {\bf 91B} (1980) 99.
  
\bibitem{inflation1} A. Guth , Phys. Rev. D \textbf{23}, 347 (1981).

\bibitem{inflation2} A. Albrecht and P. J. Steinhardt, Phys. Rev. Lett. \textbf{48}, 1220 (1982).

\bibitem{inflation3} A.~D.~Linde,
  Phys.\ Lett.\ B {\bf 129} (1983) 177.
  
\bibitem{Abazajian:2013vfg} K.~N.~Abazajian {\it et al.},
  Astropart.\ Phys.\  {\bf 63} (2015) 55
  [arXiv:1309.5381 [astro-ph.CO]].
  
\bibitem{Starobinsky:1979ty} A.~A.~Starobinsky, JETP Lett.\  {\bf 30}, 682 (1979).

\bibitem {R2}V.F. Mukhanov and G.V. Chibisov , JETP Letters \textbf{33}, 532(1981)

\bibitem {R202}S. W. Hawking,Phys. Lett. B \textbf{115}, 295 (1982)

\bibitem {R203}A. Guth and S.-Y. Pi, Phys. Rev. Lett. \textbf{49}, 1110 (1982)

\bibitem {R204}A. A. Starobinsky, Phys. Lett. B \textbf{117}, 175 (1982)

\bibitem {R205}J.M. Bardeen, P.J. Steinhardt and M.S. Turner, Phys. Rev.D
\textbf{28}, 679 (1983).

\bibitem{Lyth:2009zz}
  D.~H.~Lyth and A.~R.~Liddle,
  Cambridge, UK: Cambridge Univ. Pr. (2009) 497 p


  
\bibitem{reh1} L.~F.~Abbott, E.~Farhi and M.~B.~Wise,
  Phys.\ Lett.\  {\bf 117B} (1982) 29.

\bibitem{reh2} A.~Albrecht, P.~J.~Steinhardt, M.~S.~Turner and F.~Wilczek,
  Phys.\ Rev.\ Lett.\  {\bf 48} (1982) 1437.
  
\bibitem{paper1} L.~Dai, M.~Kamionkowski and J.~Wang,
  Phys.\ Rev.\ Lett.\  {\bf 113} (2014) 041302
  [arXiv:1404.6704 [astro-ph.CO]].

\bibitem{paper2} J.~B.~Munoz and M.~Kamionkowski,
  Phys.\ Rev.\ D {\bf 91} (2015) no.4,  043521
  [arXiv:1412.0656 [astro-ph.CO]].

\bibitem{paper3} J.~L.~Cook, E.~Dimastrogiovanni, D.~A.~Easson and L.~M.~Krauss,
  JCAP {\bf 1504} (2015) 047
  [arXiv:1502.04673 [astro-ph.CO]].
  
\bibitem{SN1} A.~G.~Riess et al. Astron. J. 116, 1009 (1998).

\bibitem{SN2} S.~Perlmutter et al., Astrophys. J. 517, 565 (1999).

\bibitem{wmap} G.~Hinshaw {\it et al.} [WMAP Collaboration],
  Astrophys.\ J.\ Suppl.\  {\bf 208} (2013) 19
  [arXiv:1212.5226 [astro-ph.CO]].
  
\bibitem{planck} P.~A.~R.~Ade {\it et al.} [Planck Collaboration],
  Astron.\ Astrophys.\  {\bf 594} (2016) A13
  [arXiv:1502.01589 [astro-ph.CO]].
  
  \bibitem{Aghanim:2018eyx}
  N.~Aghanim {\it et al.} [Planck Collaboration],
  arXiv:1807.06209 [astro-ph.CO].
  
\bibitem{BAO1} D.~J.~Eisenstein {\it et al.} [SDSS Collaboration],
  Astrophys.\ J.\  {\bf 633} (2005) 560
  [astro-ph/0501171].
  
\bibitem{BAO2} W.~J.~Percival {\it et al.} [SDSS Collaboration],
  Mon.\ Not.\ Roy.\ Astron.\ Soc.\  {\bf 401} (2010) 2148
  [arXiv:0907.1660 [astro-ph.CO]].
  
\bibitem{weinberg} S.~Weinberg,
  Rev.\ Mod.\ Phys.\  {\bf 61} (1989) 1.  

\bibitem{modified1} T.~P.~Sotiriou and V.~Faraoni,
  Rev.\ Mod.\ Phys.\  {\bf 82} (2010) 451
[arXiv:0805.1726 [gr-qc]].

\bibitem{modified2} A.~De Felice and S.~Tsujikawa,
Living Rev.\ Rel.\  {\bf 13} (2010) 3
[arXiv:1002.4928 [gr-qc]].

\bibitem{BD} C.~Brans and R.~H.~Dicke,
  Phys.\ Rev.\  {\bf 124} (1961) 925.

\bibitem{dgp} G.~R.~Dvali, G.~Gabadadze and M.~Porrati,
  Phys.\ Lett.\ B {\bf 485} (2000) 208
[hep-th/0005016].

\bibitem{galileon} A.~Nicolis, R.~Rattazzi and E.~Trincherini,
  Phys.\ Rev.\ D {\bf 79} (2009) 064036
  [arXiv:0811.2197 [hep-th]].
  
\bibitem{review} T.~Clifton, P.~G.~Ferreira, A.~Padilla and C.~Skordis,
  Phys.\ Rept.\  {\bf 513} (2012) 1
  [arXiv:1106.2476 [astro-ph.CO]].
  
\bibitem{horndeski} G.~W.~Horndeski, 
Int. \ J. \ Theor. \ Phys. 10, 363-384 (1974).  

\bibitem{defelice} A.~De Felice, T.~Kobayashi and S.~Tsujikawa,
  Phys.\ Lett.\ B {\bf 706} (2011) 123
  [arXiv:1108.4242 [gr-qc]]. 
  
\bibitem{mukhanov2} C.~Armendariz-Picon, V.~F.~Mukhanov and P.~J.~Steinhardt,
  Phys.\ Rev.\ Lett.\  {\bf 85} (2000) 4438
  [astro-ph/0004134].  
  
 
\bibitem{kobayashi} T.~Kobayashi, M.~Yamaguchi and J.~Yokoyama,
  Prog.\ Theor.\ Phys.\  {\bf 126} (2011) 511
  [arXiv:1105.5723 [hep-th]].
  
\bibitem{Kobayashi:2010cm}
  T.~Kobayashi, M.~Yamaguchi and J.~Yokoyama,
  Phys.\ Rev.\ Lett.\  {\bf 105} (2010) 231302
  [arXiv:1008.0603 [hep-th]].
  
\bibitem{deffayet} C.~Deffayet, X.~Gao, D.~A.~Steer and G.~Zahariade,
  Phys.\ Rev.\ D {\bf 84} (2011) 064039
  [arXiv:1103.3260 [hep-th]].  
  
\bibitem{charmousis} C.~Charmousis, E.~J.~Copeland, A.~Padilla and P.~M.~Saffin,
  Phys.\ Rev.\ Lett.\  {\bf 108} (2012) 051101
  [arXiv:1106.2000 [hep-th]].  
  
  
\bibitem{Kobayashi:2019hrl} 
  T.~Kobayashi,
  Rept.\ Prog.\ Phys.\  {\bf 82}, no. 8, 086901 (2019)


\bibitem{TheLIGOScientific:2017qsa}
  B.~P.~Abbott {\it et al.} [LIGO Scientific and Virgo Collaborations],
  Phys.\ Rev.\ Lett.\  {\bf 119} (2017) no.16,  161101
  arXiv:1710.05832 [gr-qc].
  
\bibitem{Monitor:2017mdv} 
  B.~P.~Abbott {\it et al.} [LIGO Scientific and Virgo and Fermi-GBM and INTEGRAL Collaborations],
  Astrophys.\ J.\  {\bf 848}, no. 2, L13 (2017)
  arXiv:1710.05834 [astro-ph.HE].
  
\bibitem{Baker:2017hug}
  T.~Baker, E.~Bellini, P.~G.~Ferreira, M.~Lagos, J.~Noller and I.~Sawicki,
  Phys.\ Rev.\ Lett.\  {\bf 119}, no. 25, 251301 (2017).

\bibitem{Langlois:2017dyl}
  D.~Langlois, R.~Saito, D.~Yamauchi and K.~Noui,
  Phys.\ Rev.\ D {\bf 97}, no. 6, 061501 (2018).


  
\bibitem{Akrami:2018odb}
  Y.~Akrami {\it et al.} [Planck Collaboration],
  arXiv:1807.06211 [astro-ph.CO].
  

\bibitem{Pich:2007vu}
  A.~Pich,
  arXiv:0705.4264 [hep-ph].

\bibitem{Higgs:1964pj}
  P.~W.~Higgs,
  Phys.\ Rev.\ Lett.\  {\bf 13} (1964) 508.
  
\bibitem{Englert:1964et}
  F.~Englert and R.~Brout,
  Phys.\ Rev.\ Lett.\  {\bf 13} (1964) 321.
  
\bibitem{Linde:1983gd}
  A.~D.~Linde,
  Phys.\ Lett.\  {\bf 129B} (1983) 177.

\bibitem{Tenkanen:2017jih}
  T.~Tenkanen,
  JCAP {\bf 1712} (2017) no.12,  001
  [arXiv:1710.02758 [astro-ph.CO]].
  
\bibitem{ref3} N.~Makino and M.~Sasaki,
  Prog.\ Theor.\ Phys.\  {\bf 86} (1991) 103.   
  
  \bibitem{Tsujikawa:2013ila}
  S.~Tsujikawa, J.~Ohashi, S.~Kuroyanagi and A.~De Felice,
  Phys.\ Rev.\ D {\bf 88} (2013) no.2,  023529
  doi:10.1103/PhysRevD.88.023529
  [arXiv:1305.3044 [astro-ph.CO]].
  
\bibitem{paperbase} J.~Ohashi and S.~Tsujikawa,
  JCAP {\bf 1210} (2012) 035
  [arXiv:1207.4879 [gr-qc]].   
 

\bibitem{Ade:2018gkx}
  P.~A.~R.~Ade {\it et al.} [BICEP2 and Keck Array Collaborations],
  Phys.\ Rev.\ Lett.\  {\bf 121} (2018) 221301
  [arXiv:1810.05216 [astro-ph.CO]].



\bibitem{Teimoori:2017jzo} 
  Z.~Teimoori and K.~Karami,
  Astrophys.\ J.\  {\bf 864}, no. 1, 41 (2018).
  
\bibitem{DeFelice:2011jm}
  A.~De Felice, S.~Tsujikawa, J.~Elliston and R.~Tavakol,
  JCAP {\bf 1108} (2011) 021
  [arXiv:1105.4685 [astro-ph.CO]].

\bibitem{Herrera:2018ker}
  R.~Herrera, N.~Videla and M.~Olivares,
  Eur.\ Phys.\ J.\ C {\bf 78} (2018) no.11,  934
  [arXiv:1806.04232 [gr-qc]].
  
\bibitem{DeFelice:2010nf}
  A.~De Felice and S.~Tsujikawa,
  Phys.\ Rev.\ D {\bf 84} (2011) 124029
  [arXiv:1008.4236 [hep-th]].
  
  \bibitem{DeFelice:2011uc}
  A.~De Felice and S.~Tsujikawa,
  Phys.\ Rev.\ D {\bf 84} (2011) 083504
  [arXiv:1107.3917 [gr-qc]].

 \bibitem{DeFelice:2013ar}
  A.~De Felice and S.~Tsujikawa,
  JCAP {\bf 1303} (2013) 030
  [arXiv:1301.5721 [hep-th]]. 
  
   
  
\bibitem{Gonzalez:2018jax}
  P.~González, G.~A.~Palma and N.~Videla,
  JCAP {\bf 1812} (2018) no.12,  001
  [arXiv:1805.10360 [hep-th]].

  
  
  
\end{thebibliography}
\end{document}